\documentclass[%
 reprint,
 superscriptaddress,
showpacs,
 amsmath,amssymb,
 aps,
 twocols,
 prl
]{revtex4-1}

\usepackage{graphicx}
\usepackage{dcolumn}
\usepackage{bm}
\usepackage{color}
\usepackage{braket}

\begin{document}
\title{Advantages and challenges in coupling an ideal gas to atomistic models in adaptive resolution simulations}

\author{Karsten Kreis}
\email[]{kreis@mpip-mainz.mpg.de}
\affiliation{Max Planck Institute for Polymer Research, Ackermannweg 10, 55128 Mainz, Germany}
\affiliation{Graduate School Materials Science in Mainz, Staudinger Weg 9, 55128 Mainz, Germany}
\author{Aoife C. Fogarty}
\email[]{fogarty@mpip-mainz.mpg.de}
\author{Kurt Kremer}
\email[]{kremer@mpip-mainz.mpg.de}
\author{Raffaello Potestio}
\email[]{potestio@mpip-mainz.mpg.de}
\affiliation{Max Planck Institute for Polymer Research, Ackermannweg 10, 55128 Mainz, Germany}

\date{\today}
\begin{abstract}
In adaptive resolution simulations, molecular fluids are modeled employing different levels of resolution in different subregions of the system. When traveling from one region to the other, particles change their resolution on the fly. One of the main advantages of such approaches is the computational efficiency gained in the coarse-grained region. In this respect the best coarse-grained system to employ in the low resolution region would be the ideal gas, making intermolecular force calculations in the coarse-grained subdomain redundant. In this case, however, a smooth coupling is challenging due to the high energetic imbalance between typical liquids and a system of non-interacting particles. In the present work, we investigate this approach, using as a test case the most biologically relevant fluid, water. We demonstrate that a successful coupling of water to the ideal gas can be achieved with current adaptive resolution methods, and discuss the issues that remain to be addressed.
\end{abstract} 
\maketitle
\section{Introduction} \label{intro}

Many soft matter systems, ranging from simple liquids to complex polymer mixtures \cite{kremer_grest,kremer_grest2,yelash_2006,muellerplathe2007} and biomolecules such as proteins \cite{McCammon:1977:Nature:895880,Karplus:1979:Nature:763343,Raiteri:2006:J-Phys-Chem-B:16494409,Lou_J-Phys-Chem-B_2006_2,Arora:2007:Proc-Natl-Acad-Sci-U-S-A:18000050,Pontiggia:2008:Biophys-J:18931260,tirion93,tir96,bah97,Micheletti:2004:Proteins:15103627,potestio2009,globisch2013}, feature a nontrivial interplay of characteristic length and time scales. Because of this property, structural or energetic changes occurring at a given scale have repercussions on others. Hence, a realistic modeling of these systems has to take into account all fine-grained details that might affect, or be affected by, larger scale features.

In most cases, however, the smallest meaningful size of the system is too large to allow its simulation with a highly detailed model. At the same time, coarse-grained models, which proved to be extremely successful in understanding the properties of many soft matter systems \cite{kremer2000,kremer_mplathe_2001,vdvegt2009,Hijon2010,noid_chapter,Noid:2013}, cannot be employed in those cases in which the fine-grained detail plays a major role.

A solution to this problem is sometimes offered by dual resolution simulations \cite{jcp,adress2,adress3,annurev,adresstoluene,adolfoprl,potestio,ensing,prlcomment2011,FritschPRL,hadress,MC_hadress,JCPagarwal2014,kreisEPL2014,JCPhadress2014}. Specifically, we consider here those cases in which the finest level of detail is needed only in a relatively small functionally or physically relevant part of the system. At the same time, the remainder is required to provide the aforementioned relevant part with the necessary thermodynamic support, namely the exchange of energy and matter, albeit not being interesting {\it per se}. This is, for example, the case of a biological system, such as a protein, immersed in a solution of water and cosolvent, whose solvation properties are correctly reproduced only if the number of solvent/cosolvent molecules is large enough to mimic the Grand Canonical limit \cite{Mukherjee:2012,Mukherjee:2013}. In this situation, the (co)solvent is not the interesting part of the computer experiment; however, it forms its largest part. In such cases one can describe the relevant subsystem with the necessary accuracy, and represent the remainder with a simpler, computationally efficient model. Adaptive resolution simulation schemes have been developed to allow this spatially inhomogeneous treatment of the system. In these approaches, the model used to describe a given component is not fixed for the entire duration of the simulation; rather, molecules freely diffuse across the domain, and their resolution, i.e. the model accuracy, is determined by the specific position they occupy in space at a given point in time.

In most cases, one tries to preserve, in the coarse-grained model, certain fundamental properties of the higher resolution system. An example is given by the structure-based coarse-graining procedures applied to molecular fluids, such as iterative Boltzmann inversion (IBI) \cite{REIT03.1}, which aims to obtain an effective potential energy reproducing the radial distribution function (RDF) of the fluid. Other choices are possible, based on the desired target properties.

However, if the scope of the simulation does not focus on the low resolution part of the system, the choice of the coarse-grained interaction may be driven mainly by system simplicity and computational efficiency, provided that the correct thermodynamical conditions are preserved in the high resolution domain. In this case, it is easy to see that the optimal choice is to have no interactions in the low resolution domain. In practical terms, this means coupling the system in the high resolution region to an ideal gas. Besides computational efficiency, such an approach has several other advantages. One is, for example, the faster diffusion in the low resolution domain, which would accelerate the mixing of solvent molecules in different parts of the system, thereby improving the sampling. Another advantage is the possibility of varying the number of particles in the system at will by simply creating or destroying molecules in a region of the system where they are all ``invisible'' to each other. This would enable the simulation of a system with fixed chemical potential rather than number of particles, that is, to simulate a truly Grand Canonical ensemble with minimal computational effort.

The present work is devoted to the investigation of the practical feasibility of the coupling of the most biologically relevant fluid, namely water, to an ideal gas of point-like, non-interacting particles each of which corresponds to a water molecule. Two different but related simulation approaches, the force-based adaptive resolution simulation (AdResS) \cite{jcp,adress2,adress3,annurev} scheme and the energy-based Hamiltonian AdResS (H-AdResS) \cite{hadress,MC_hadress,kreisEPL2014,JCPhadress2014} scheme are employed to perform this coupling, and various strategies are tested and compared to preserve the correct thermodynamics in the high resolution region, where the system is a fully atomistic fluid.

The most immediate advantage of replacing the vast majority of the solvent molecules in the system with an ideal gas is clearly the lack of any force calculation in the low resolution domain. This benefit naturally comes at some cost, namely the large thermodynamical differences existing between the equations of state of an ideal gas and a generic molecular fluid. These free energy discrepancies nontrivially affect the behavior of the hybrid, dual resolution system, and require particular care in the construction of the interface, in order to allow the atomistic, high resolution region to behave as expected. Several strategies have been developed in the past to modulate the thermodynamical balance between the two coupled models \cite{FritschPRL,hadress}, but these were in most cases applied to a ``well behaved'' coarse-grained system, parametrized for the corresponding atomistic system. The large free energy difference, as well as diffusion dynamics that differ by orders of magnitude, make the construction of a smooth seam between water and ideal gas a challenging problem.

\begin{figure}
	\includegraphics[width=\columnwidth]{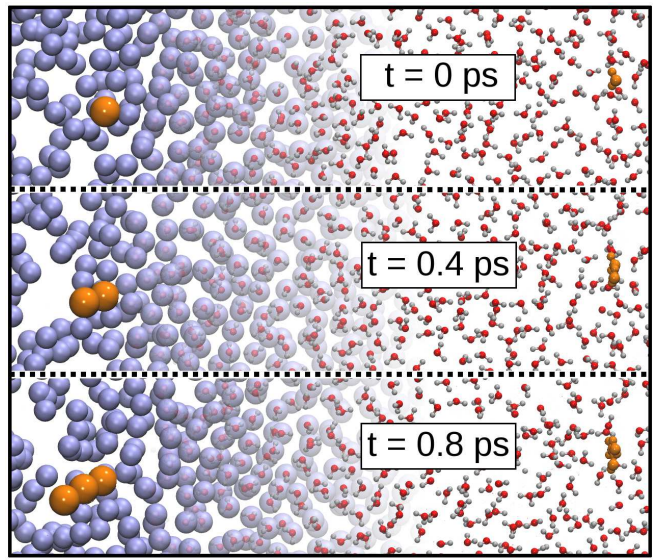} 
\caption{Three consecutive snapshots of the water--ideal gas interface. For clarity we show only molecules in a $0.7$ nm thick layer in the direction perpendicular to the sheet. The water, in the right half of the figure, is much more structured than the ideal gas (left), in which several molecules overlap. For the orange highlighted molecules the time evolution is visualized by copying their previous positions into the subsequent snapshots. From this it can be seen that the molecules in the ideal gas diffuse much faster than the water molecules, whose positions almost do not change over $0.8$ ps.}
\label{fig:visual}       
\end{figure}

\section{Adaptive resolution simulations} \label{adres}

As briefly sketched in the Introduction, the idea lying at the core of dual resolution simulations is to introduce into the simulation domain of a soft matter system a spatial, geometrical separation between two of its parts. One part, typically the smallest, needs to be described with a computationally expensive, high resolution model, and will be henceforth referred to as the atomistic (AT) region, assuming that the single-atom level is the finest we can and need to reach. (This is not the general case, however, and even finer models can be employed, as e.g. in \cite{adolfoprl,potestio}.) The other, larger part of the system is necessary inasmuch as it represents at least the thermal bath and particle reservoir of the small subsystem, but is not interesting in itself and can therefore be modeled in terms of coarse-grained (CG) particles and force fields. This subpart retains simpler, smoother non-bonded interactions, which require a smaller number of force calculations.

These two domains are joined together through a hybrid (HY) region, by which one set of interactions is gradually transformed into the other. This smooth change is parametrized in terms of a resolution, or switching, function $\lambda(x)$, which continuously and monotonically goes from $1$ (in the AT region) to $0$ (in the CG region). In the HY region the interactions are a combination of the atomistic and coarse-grained force fields, the specific form of this interpolation being specific to a given adaptive resolution simulation scheme. Molecules are free to diffuse throughout the whole simulation domain, and their resolution changes dynamically according to their instantaneous position in space.

In general, it is necessary to take into account the fact that the two models follow different equations of state. Therefore the equilibrium state they attain once coupled, each in its pertinent domain, is typically not the state they would have attained if they occupied the entire simulation box \cite{FritschPRL,hadress}. Hence, system-specific modifications have to be enforced to modulate the local equilibrium towards the desired state.

The following subsections \ref{FbAdress} and \ref{EbAdress} are devoted to the description of the two dual resolution simulation schemes employed in the present work. The main difference between them lies in the interpolation of the interactions in the HY interface: the first one, AdResS, is built based on a linear combination of forces; the H-AdResS scheme is formulated, as the name suggests, in terms of a Hamiltonian, and the interpolation of the interactions occurs first at the level of potential energies.

\subsection{Force-based approach} \label{FbAdress}

The adaptive resolution simulation (AdResS) \cite{jcp,adress2,adress3,annurev} scheme is based on the direct interpolation of two force fields. This approach satisfies Newton's Third Law exactly and instantaneously in every part of the system, including the HY region. The atomistic and coarse-grained forces acting between two molecules are linearly interpolated in a symmetric fashion, as follows:

\begin{eqnarray}\label{adress_f} \nonumber
\textbf{F}_{\alpha \beta} &=& \lambda({\bf R}_\alpha) \lambda({\bf R}_\beta) \textbf{F}^{AT}_{\alpha \beta}\\
&+& \left(1 - \lambda({\bf R}_\alpha) \lambda({\bf R}_\beta) \right) \textbf{F}^{CG}_{\alpha \beta}
\end{eqnarray}

In Eq. \ref{adress_f}, ${\bf R}_\alpha$ (resp. ${\bf R}_\beta$) is the centre of mass coordinate of molecule $\alpha$ (resp. $\beta$).
$\textbf{F}^{AT}_{\alpha \beta}$ and $\textbf{F}^{CG}_{\alpha \beta}$ are, respectively, the atomistic and the coarse-grained forces acting on molecule $\alpha$ due to the interaction with molecule $\beta$.

The coupling of two different models of the same system naturally leads to a thermodynamical imbalance, e.g. if one of the two has, for a given temperature and density, a higher virial pressure than the other. In this particular case, the system will evolve towards an equilibrium state in which the pressure gradients are flattened out, but the density profile of the system will not be uniform in the direction of resolution change. To enforce a uniform density profile, one can make use of an external field, called Thermodynamic Force \cite{FritschPRL} (TF), which is obtained iteratively according to the following update scheme:

\begin{eqnarray}\label{thermoforce}
{\bf f}_{th}^{i+1} = {\bf f}_{th}^i - \frac{M}{\rho_0^2 \kappa_T}\nabla \rho^i(r)
\end{eqnarray}

\noindent
where $\rho^i(r)$ is the density profile as a function of position after the $i$-th iteration, $M$ is the molecular mass, $\rho_0$ is the reference density and $\kappa_T$ is the isothermal compressibility of the fluid. By construction, this iterative protocol reaches a fixed point when the density is uniform, and the update term $\nabla \rho$ is zero. It is worth mentioning that enforcing a flat density profile is not the only option: in fact, one might wish to keep other thermodynamical quantities, such as pressure, or higher-order correlations, e.g. RDFs, constant throughout the system. In principle one can enforce, in the whole system, a uniform profile for two or more of these quantities, but this possibility ultimately depends on the coarse-grained model: in some cases, in fact, having a CG force field that correctly reproduces the reference value of a given thermodynamic quantity implies the impossibility to do the same for one other quantity which is conjugate to the first. A well-known example of this is provided by the pressure and the compressibility of a CG potential obtained {\it via} IBI, which cannot be simultaneously matched \cite{wang2009comparative}. On the other hand, this limitation of the CG force field does not turn into a limitation of the adaptive approach, since in most cases the only requirement is to have the desired thermodynamics correctly reproduced in the sole AT region.

The force field in Eq. \ref{adress_f} is intrinsically non-conservative, as it cannot be obtained as the negative gradient of a potential energy function \cite{prelu}. This fact determines some limitations to the applicability of this approach, most notably the inability to perform Microcanonical and Monte Carlo simulations. Additionally, a local thermostat is required to enforce a state of dynamical equilibrium, in which the temperature is constant. On the other hand, it can be proven that the system correctly samples the Canonical ensemble in the AT region, and the configurations generated in a molecular dynamics simulation with the AdResS scheme are compatible with thermal equilibrium \cite{dellesite_prx,JCPagarwal2014}.

\subsection{Energy-based approach} \label{EbAdress}

The second simulation strategy we employ in the present work is the Hamiltonian AdResS (H-AdResS) \cite{hadress,MC_hadress,kreisEPL2014,JCPhadress2014} method, in which two models of a system are coupled directly at the level of potential energies. The H-AdResS scheme is hence formulated in terms of a potential energy function, defined as:

\begin{eqnarray}\label{hadress_H}
V_{H-AdResS} &=&\\ \nonumber
&=&\mathcal K + V^{int} + \sum_{\alpha}^N \left\{{\lambda_\alpha} {V^{AT}_\alpha} + {(1 - \lambda_\alpha)} {V^{CG}_\alpha} \right\}
\end{eqnarray}
where $N$ is the number of molecules, $\mathcal K$ is the kinetic energy, $V^{int}$ is the internal potential energy of the molecules, and:
\[
 \left\{
  \begin{array}{l}
{V^{{AT}}_\alpha = \displaystyle\frac{1}{2}\sum_{\beta,\beta\neq \alpha}^{N} \sum_{ij} V^{AT}(|\mbox{\bf r}_{\alpha i} - \mbox{\bf r}_{\beta j}|)}\\
{V^{{CG}}_\alpha = \displaystyle\frac{1}{2}\sum_{\beta,\beta\neq \alpha}^{N} V^{CG}(|\mbox{\bf R}_\alpha - \mbox{\bf R}_\beta|)}\\
{\lambda_\alpha = \lambda(\mbox{\bf R}_\alpha)}\\
  \end{array} \right.\vspace{6pt}
\]

Contrary to the force-based approach, H-AdResS also allows NVE and MC simulations. The drawback is the presence in the force field of a term proportional to the gradient of the switching function. In fact, the force acting on molecule $\alpha$ reads:
\begin{eqnarray}\label{hadress_F} \nonumber
\textbf{F}_{\alpha} =&& \sum_{\beta,\beta\neq \alpha} \left\{ \frac{\lambda_\alpha + \lambda_\beta}{2} \textbf{F}^{AT}_{\alpha |\beta} + \left(1 - \frac{\lambda_\alpha + \lambda_\beta}{2}\right) \textbf{F}^{CG}_{\alpha |\beta} \right\}\\
&&- \left[ V^{AT}_\alpha - V^{CG}_\alpha \right] \nabla_{\alpha}\lambda_\alpha
\end{eqnarray}

The first term of Eq. \ref{hadress_F} bears some similarity with the AdResS force interpolation of Eq. \ref{adress_f}: both, in fact, are anti-symmetric with respect to molecule label exchange. On the other hand, they differ already at the level of the interpolation weights: in the AdResS case, in fact, they are given by the product of the switching functions of the two molecules, while in the H-AdResS case the average of the $\lambda$'s naturally appears. This difference mainly results in diverse interactions between molecules in the CG region and those in the hybrid interface. The AdResS scheme treats the interaction with a molecule in the CG domain as purely CG, while in the H-AdResS scheme the CG molecules interact also at the atomistic level with the ones in the hybrid region, even though only weakly. For all practical purposes, however, this difference bears no consequences.

The last term contains the largest difference with the force-based method, namely a term proportional to the gradient of the switching function. This term, referred to as the drift force, contains the difference between the atomistic and the coarse-grained potential energy of a molecule, and is zero outside the HY region, where $\nabla\lambda \equiv 0$ by construction. In the HY region, though, it locally breaks Newton's Third Law, inasmuch as the force acting between molecule pairs cannot be written as a sum of antisymmetric terms.

The drift force can be compensated on average by including in the definition of the Hamiltonian a new term conceptually similar to the thermodynamic force employed in the AdResS scheme \cite{hadress,MC_hadress}:

\begin{eqnarray}\label{eq:fec:01}
V_{\Delta} = V_{H-AdResS} - \sum_{\alpha=1}^N \Delta H(\lambda(\textbf{R}_\alpha))
\end{eqnarray}

The functional form of this term is defined by the requirement that its corresponding force and the drift force cancel out, i.e.:

\begin{eqnarray}
&&\frac{d \Delta H(\lambda)}{d \lambda}\biggl|_{\lambda = \lambda_\alpha} = \left\langle  \left[ V^{AT}_\alpha - V^{CG}_\alpha \right] \right\rangle_{{\bf R}_\alpha}
\end{eqnarray}
where the subscript in the average indicates that the latter has to be performed constraining the coarse-grained site of molecule $\alpha$ in the position ${\bf R}_\alpha$. In many cases the term $\Delta H$ can be well approximated by the Helmholtz free energy difference between the coarse-grained model and a hybrid model with mixing parameter $\lambda$. In practice $\Delta H$ can be obtained by means of a Kirkwood Thermodynamic Integration \cite{kirkwood1935,hadress,MC_hadress} (TI):
\begin{eqnarray}\label{kirk1}\nonumber
\Delta H(\lambda) &=& \int_0^\lambda d\lambda' \frac{d \Delta H(\lambda')}{d \lambda'}\\  \nonumber
&\simeq& \frac{1}{N} \int_0^{\lambda} d\lambda' \left\langle \left[ V^{AT} - V^{CG} \right]  \right\rangle_{\lambda'}\\
&=&\frac{\Delta F(\lambda)}{N}
\end{eqnarray}

This choice of $\Delta H$ removes the average effect of the drift force, thereby restoring, although only on average, Newton's Third Law also in the HY region. However, it does not guarantee that the density in the AT and the CG domain attains the same value. To this end, it is necessary to add a term proportional to the pressure difference between the two models, which amounts to choosing, for $\Delta H$, the chemical potential difference as a function of the resolution: 

\begin{eqnarray}
\Delta H(\lambda) = \frac{\Delta F(\lambda)}{N} + \frac{\Delta p(\lambda)}{\rho_0} \equiv \Delta \mu(\lambda) 
\end{eqnarray}

This correction to the Hamiltonian takes the name of free energy compensation (FEC) \cite{hadress,MC_hadress}.

\section{Simulation details}\label{sec:methodo}

All simulations presented here used a system containing 6526 particles (water molecules and ideal gas particles) in a simulation box of dimensions $\sim 16.1 \times 3.5 \times 3.5$ nm. This corresponds to a density of 33.1 molecules nm$^{-3}$ or 990.7 kg m$^{-3}$, a value determined via fully atomistic simulations at a pressure of 1~bar and 300~K in the isothermal-isobaric ensemble, and close to the experimental density under ambient conditions. The instantaneous resolution of a given particle was determined by the distance $d_x$ along the $X$-axis between its centre of mass and the centre of the simulation box. The atomistic region was defined as $d_x \leq 3.0$~nm, flanked by two HY regions at $3.0 < d_x < 5.0$~nm, and the coarse-grained region at $d_x \geq 5.0$~nm. Periodic boundary conditions were used. The water--ideal gas interface region is illustrated in Fig. \ref{fig:visual}. To assign to a molecule its position-dependent resolution $\lambda$, its distance from the boundary between the atomistic and the HY region is computed, i.e. $d_x-d_{\text{at}}/2$, where $d_{\text{at}}$ is the width of the atomistic region. This quantity is then inserted into the resolution function $\lambda(x)$, which is given as:
\begin{equation}\label{ResFct}
\lambda(x) =
\left\{
	\begin{array}{ll}
		1 & \mbox{: } x \leq 0 \\
		1-\frac{30}{d_{\text{hy}}^5}(\frac{1}{5}x^5-\frac{d_{\text{hy}}}{2}x^4+\frac{d_{\text{hy}}^2}{3}x^3) & \mbox{: } 0 < x < d_{\text{hy}} \\
		0 & \mbox{: } x \geq d_{\text{hy}}
	\end{array}
\right.
\end{equation}
where $d_{\text{hy}}$ denotes the width of the HY region. In our simulations, $d_{at} = 6$ nm and $d_{hy} = 2$ nm.

Simulations were performed using the ESPResSo++ package \cite{espressopp}, and a timestep of 2~fs. Atomistic water was represented using the SPC/E model \cite{Berendsen_jphyschem_1987-SPCE} and the SETTLE algorithm \cite{Kollman_JCompChem_1992-SETTLE}. Non-bonded interactions used a cutoff of 1.0~nm. Electrostatic interactions were calculated using the reaction field method with a dielectric constant of 67.5998, as previously determined for SPC/E water \cite{delleSite_JCTC_2012-KBliqmix}. Production runs had a length of between 200~ps and 2~ns, depending on the convergence time of the properties studied, and were performed in the canonical ensemble at a temperature of 300~K, using the Langevin thermostat with a friction constant of 0.5~ps$^{-1}$ in H-AdResS and 5.0~ps$^{-1}$ in AdResS. The stronger coupling to the thermostat in the force-based scheme is necessary in order to counteract the excess heat produced in the HY region due to the removal of degrees of freedom and the non-conservative nature of the force interpolation simulations \cite{JCPagarwal2014}.
The use of non-interacting ideal gas particles in the coarse-grained region allows particle positions to overlap, and two overlapping ideal-gas particles entering the HY region, where they begin to interact, may lead to unmanageably large forces. The absolute magnitude of the force between any particle pair was therefore capped at $10^4$ kJ mol$^{-1}$ nm$^{-1}$, in order to allow particle pairs to adapt their inter-particle distance as interaction strength gradually increases across the HY region.

As well as coupling the atomistic water model to an ideal gas, for comparison we also performed simulations coupling atomistic water to a coarse-grained potential developed using Iterative Boltzmann Inversion with pressure correction \cite{REIT03.1}, which reproduces the pressure of the underlying atomistic model and provides an excellent approximation of its RDF. This IBI potential acted on particle centres of mass, and was obtained using the VOTCA coarse-graining package \cite{Andrienko_JCTC_2009-votca}, and 300 IBI steps, each running for 100 ps.

\section{Results and discussion} \label{results}

\subsection{Force-based coupling with the AdResS method} \label{results:adress}

First, we couple the atomistic water model to an ideal gas using the force-based AdResS scheme. In this case, i.e. when the coarse-grained interaction is absent, Eq. \ref{adress_f} reduces to:

\begin{eqnarray}\label{adress_f_IG}
\textbf{F}_{\alpha \beta} = \lambda({\bf R}_\alpha) \lambda({\bf R}_\beta) \textbf{F}^{AT}_{\alpha \beta}\end{eqnarray}

\begin{figure}
	\includegraphics[width=\columnwidth]{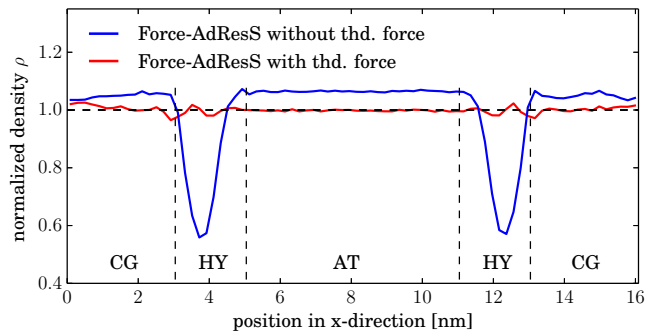} 
\caption{Density profiles for force-AdResS water--ideal gas system, with and without thermodynamic force.}
\label{fig:dens-prof-fadres}
\end{figure}

A direct coupling of the liquid to the ideal gas produces the density profile reported in Fig. \ref{fig:dens-prof-fadres} (blue line). A considerable depletion can be observed in the HY region, where the density drops to less than $60$\% of the reference value. Correspondingly, in the AT and CG region the density is higher, but the value attained is almost the same in both domains. The marked dip in the density profile is not symmetric and is located closer to the border with the CG than with the AT region. Its origin can be understood by considering that particles entering the HY region from the CG region, where they were non-interacting, may be located very close together in space. In order to permit stable simulations this is dealt with using force-capping, as outlined above; the high forces nonetheless lead to a peak in the virial pressure profile at the point corresponding to the dip in the density profile, when an AdResS simulation is launched from an initial configuration with a uniform density profile.

\begin{figure}
	\includegraphics[width=\columnwidth]{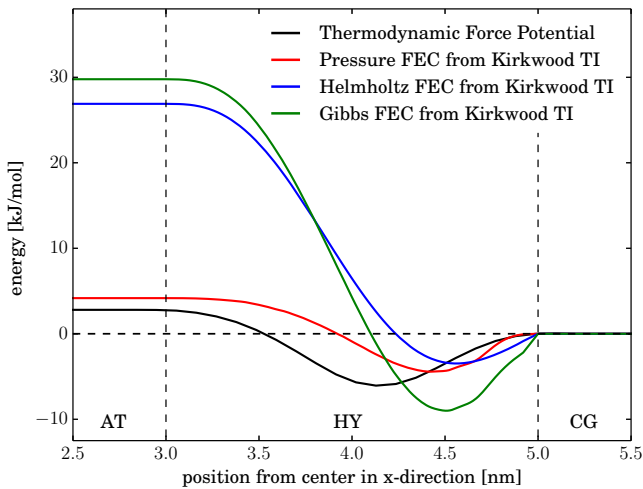} 
\caption{Potential energies corresponding to the Thermodynamic Force and FEC contributions as obtained from Kirkwood TI. The Gibbs FEC is the sum of the Helmholtz and the pressure FECs.}
\label{fig:pressurepotentials}       
\end{figure}

In order to remove this depletion we make use of the TF iterative correction, Eq. \ref{thermoforce}, discussed in Sect. \ref{FbAdress}. The potential energy function corresponding to the converged TF force field is reported in Fig. \ref{fig:pressurepotentials}. As shown in Fig. \ref{fig:dens-prof-fadres} (red line), the application of the TF enforces a perfectly flat density profile in the AT region, attaining the reference value. Small fluctuations, of the order of $1-2$\%, can be observed in the HY region. In the CG region the ideal gas density deviates by almost the same amount from the reference. The slightly noisy profile in the latter region is due to the difficulty to average the position of non-interacting particles.

\subsection{Energy-based coupling with the H-AdResS method} \label{results:Hadress}

Next, we move to the case of the H-AdResS method. In the ideal gas case, the H-AdResS Hamiltonian (without FEC) becomes:
\begin{eqnarray}\label{hadress_H_ig}
&&H = \mathcal K + V^{int} + \sum_{\alpha} {\lambda_\alpha} {V^{AT}_\alpha}
\end{eqnarray}

\begin{figure}
  \includegraphics[width=\columnwidth]{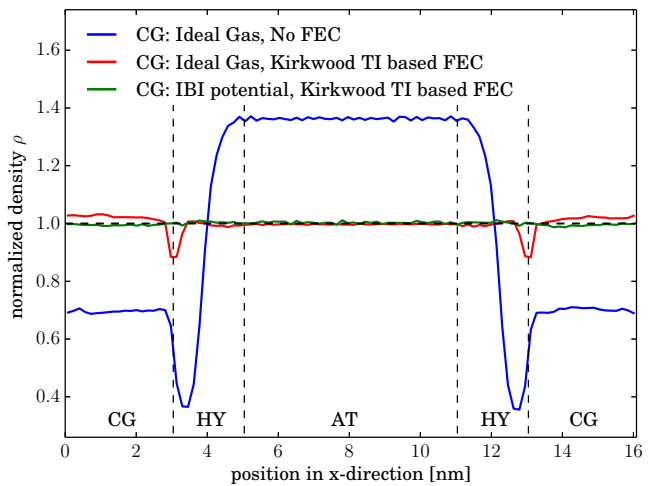} 
\caption{Density profiles for H-AdResS simulations of the water--ideal gas system with and without Kirkwood TI based Gibbs FEC as well as for the water--IBI system, also with Kirkwood TI based Gibbs FEC.}
\label{fig:2}       
\end{figure}

In this case, as in the AdResS case, the simple coupling of atomistic water to the ideal gas results in a depletion of molecules in the HY region, see Fig. \ref{fig:2}. Additionally, we observe a different density between the AT and the CG regions, due to the presence of the drift force. Applying the FEC, whose plot is reported in Fig. \ref{fig:pressurepotentials}, we see that the density attains the reference value in the whole AT region and in most of the HY region. A small depletion of approximately $10$\% is observed at the HY/CG interface, then the density in the CG region flattens out again, to a value $1-2$\% higher than the reference. Also in this case the higher density in the CG domain is due to the depletion in the HY region. As already mentioned, the origin of this depletion lies in overlapping pairs of molecules diffusing from the CG region into the HY layer, where their high intermolecular forces lead to high virial pressure. The deviations from the reference density observed in the hybrid simulations making use of the ideal gas as a coarse-grained model are relatively small, do not affect the AT region and, as it will be discussed below, the density dip does not prevent molecules from diffusing across the interfaces separating different resolutions. The application of the thermodynamic force scheme already employed in the AdResS case would in any case remove any minor deviation from the reference density.

For comparison, we performed a simulation of the same atomistic water model coupled to a coarse-grained model obtained {\it via} iterative pressure-corrected IBI. In this case, the well-parametrized coarse-grained model, together with the FEC field, enforces a very uniform density throughout the system. This demonstrates that when the fully atomistic system is coupled to a coarse-grained model reproducing at least a few thermodynamical properties of the former at the same state point, the FEC term is sufficiently accurate to remove the remaining discrepancies in the free energy. In the ideal gas case, in contrast, the mean-field character of the FEC becomes apparent in the imperfect correction occurring at the HY/CG interface, where local correlations are present that are not compensated for. The trade-off between a non-flat density profile and the advantage of not having to parametrize the coarse-grained model depends on the specific system under consideration.

\begin{figure}
  \includegraphics[width=\columnwidth]{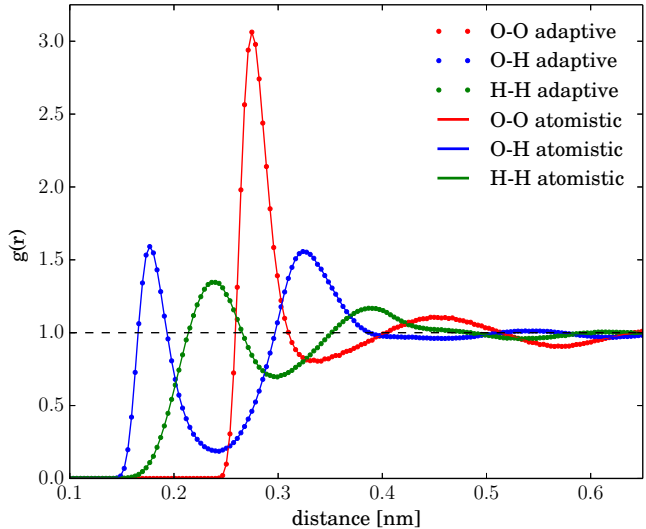} 
\caption{RDFs for pure water and for the atomistic region of the H-AdResS water--ideal gas with Kirkwood TI based Gibbs FEC. Since a rigid water model is employed, all RDFs consider only intermolecular atom pairs.}
\label{fig:3}       
\end{figure}

We now move on to the quantitative assessment of the correctness of the structure of the fluid in the AT region of the water--ideal gas hybrid system. We measured the pair correlation functions for pairs of atoms in which at least one atom had a distance along the $X$-axis of less than 1.5~nm from the centre of the atomistic region. Since the sum of this distance and the RDF cutoff is less than the distance used to define the atomistic region, the RDFs calculated are fully atomistic. These RDFs are fully consistent with those measured in completely atomistic reference simulations, as is evident in Fig. \ref{fig:3}. 

A crucial point is to verify that the system in the AT region behaves as if it were a subpart of a completely atomistic system. This means not only measuring the density profile and the RDFs, but also checking that the molecules do not have any impediment in diffusing across the HY region. To this end, we followed the time evolution of a subset of labelled particles at the AT/HY and CG/HY interface, to check that nothing prevents them from moving across the system. Obviously, we can expect a different diffusion rate in the two cases, as the friction of the ideal gas is decidedly smaller than that experienced by the atomistic water molecules.

\begin{figure}
	\includegraphics[width=\columnwidth]{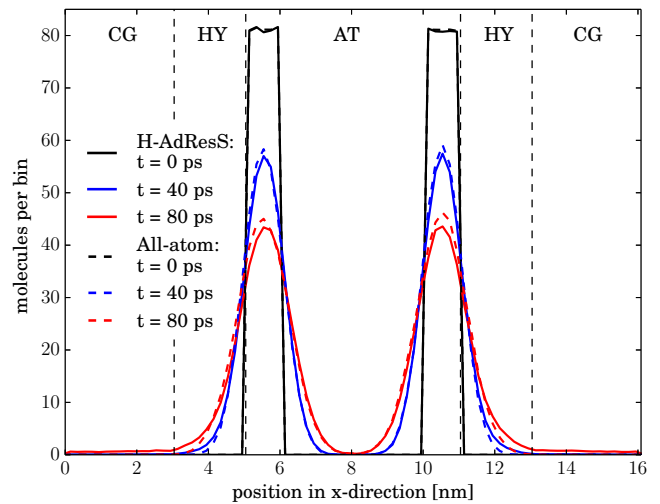} 
\caption{Diffusion profiles in H-AdResS simulations of the water-ideal gas system and in fully atomistic reference simulations of SPC/E water: time evolution of the position of molecules initially located in a 1-nm-wide slab in the atomistic region, immediately adjacent to the HY region. The y-axis is the absolute number of these molecules whose centre of mass $X$-coordinate is in a given bin at the given time.}
\label{fig:diff-prof-hadres-vs-aa}       
\end{figure}

In Fig. \ref{fig:diff-prof-hadres-vs-aa} (solid lines) we report the diffusion profiles of the molecules initially located in two symmetric slabs of width 1~nm in the AT region, at the interfaces with the HY regions. These molecules, uniformly distributed at $t=0$, spread out throughout the whole system as time passes. The overall distribution is quasi-Gaussian, but the half moving towards the CG region extends further than the half in the AT region, as the friction in the former allows a faster diffusion.

The diffusion in the AT region, though, is perfectly compatible with that of a fully atomistic water system. This can be verified by comparing the diffusion profiles of the H-AdResS simulation with those obtained by performing the same analysis on a fully atomistic system. These latter distributions, reported in Fig. \ref{fig:diff-prof-hadres-vs-aa} (dashed lines), overlap very well in the AT region and even in part of the HY region, while, as expected, the distribution of the molecules in the H-AdResS simulation (solid lines) extends deeper in the CG region. This asymmetry is due to the fact that the CG model, the ideal gas, has by construction very different transport properties with respect to the atomistic water model, and differences between the two cases have to be present. Our attention, though, focusses on the atomistic subdomain, where the diffusion occurs exactly as in the reference system, and this is the only aspect we deem crucial.

\begin{figure}
	\includegraphics[width=\columnwidth]{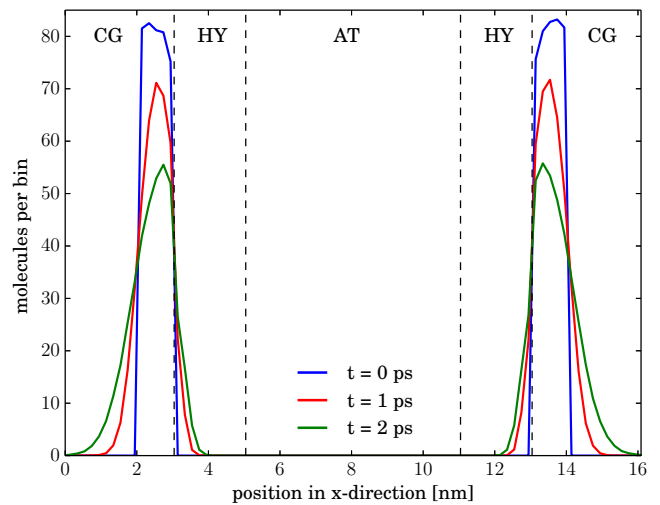} 
\caption{Diffusion profiles in H-AdResS simulations of the SPC/E water-ideal gas system: time evolution of the position of molecules initially located in a 1-nm-wide slab in the coarse-grained region, immediately adjacent to the HY region. The y-axis is the absolute number of these molecules whose centre of mass $X$-coordinate is in a given bin at the given time.}
\label{fig:diff-prof-hadres-cgregion}       
\end{figure}

This difference between the equilibration times of atomistic water and of the ideal gas is most clearly seen when comparing the plots in Fig. \ref{fig:diff-prof-hadres-vs-aa} (solid lines) with those showing the diffusion of molecules at the CG/HY interface, Fig. \ref{fig:diff-prof-hadres-cgregion}. Not only are the distributions for $t>0$ more skewed than the ones previously shown, they also evolve on a  much faster time scale. For the distribution peak to reach the height that the particles at the AT/HY interface attain at $t=40$ ps, those in the CG/HY interface need only $2$ ps.

\begin{figure}
	\includegraphics[width=\columnwidth]{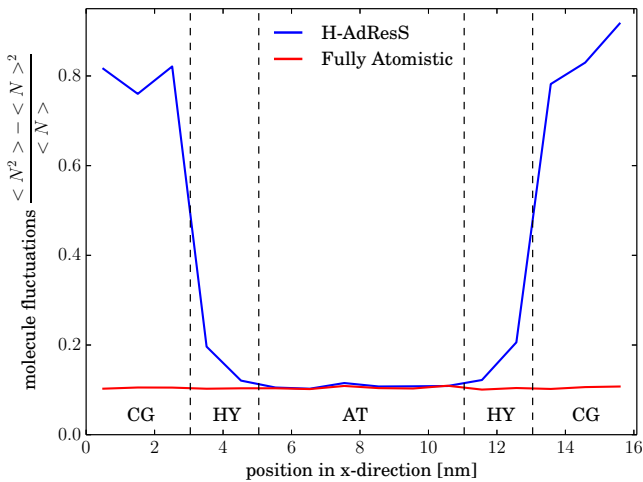} 
\caption{Molecule fluctuations as a function of position in the simulation box for H-AdResS water--ideal gas simulation with Gibbs FEC as well as for fully atomistic reference simulations.}
\label{fig:fluctuations}       
\end{figure}

Finally, we studied the density fluctuations across the system, since these can be expected to differ enormously between a fluid of non-interacting particles and a condensed, strongly interacting fluid. We measured the molecule number fluctuations, a quantity proportional to the compressibility and defined as
\begin{equation}
\Delta N \equiv \frac{\langle N^2\rangle-\langle N\rangle^2}{\langle N\rangle}
\end{equation}
where $N$ is the number of particles in a 1-nm wide slab of the simulation box. The resulting profiles are shown in Fig.~\ref{fig:fluctuations}. Although the local density fluctuations in the ideal gas region are almost an order of magnitude higher than those in the atomistic region, the latter nonetheless correspond perfectly to the local density fluctuations in a fully atomistic system.

\subsection{Computational gain} \label{results:speedup}

When performing adaptive dual resolution simulations, one of the main advantages when coupling to a coarse-grained potential is the computational gain over detailed, fully atomistic simulations. In this respect the ideal gas is the most efficient coarse-grained model, as it corresponds to no interaction at all. In this section we report a comparison of the performance of the adaptive water--ideal gas system and a fully atomistic water setup. Additionally, we also discuss the ideal gas compared to other coarse-grained models, i.e. the coarse-grained potential from IBI.

We performed 4 sets of simulations, with four different box lengths, each consisting of an atomistic simulation as well as a water--IBI and a water--ideal gas H-AdResS simulation. Each of them ran for $10$ ps. In the H-AdResS simulations, Kirkwood TI based Gibbs FEC's were employed in order to keep the densities flat (compare Fig. \ref{fig:2}). In all cases under examination the atomistic region has a total width of $2.0\,\text{nm}$; similarly, the adjacent HY regions always have widths of $2.0\,\text{nm}$ each. The size of the CG region, however, was chosen differently for all sets; the total simulation box sizes are presented, together with the corresponding molecule numbers, in Tab. \ref{tab_speedup}. In applications of adaptive resolution schemes, the high resolution region is typically, though not necessarily, much smaller than the coarse-grained region. Therefore, in our adaptive test setups the AT and HY regions occupy only a relatively small volume in the simulation boxes compared to the CG part. 

\begin{table}[!ht]\renewcommand{\arraystretch}{1.3}\addtolength{\tabcolsep}{-1pt}\renewcommand{\tabcolsep}{0.3cm}
\begin{center}
\begin{tabular}{c c c c}
\hline
\# molecules & $L_x$ & $L_y$ & $L_z$  \\ 
\hline
6526 & $16.086\,\text{nm}$ & $3.500\,\text{nm}$ & $3.500\,\text{nm}$ \\
9803 & $24.164\,\text{nm}$ & $3.500\,\text{nm}$ & $3.500\,\text{nm}$ \\
13064 & $32.202\,\text{nm}$ & $3.500\,\text{nm}$ & $3.500\,\text{nm}$ \\
16349 & $40.300\,\text{nm}$ & $3.500\,\text{nm}$ & $3.500\,\text{nm}$ \\
\hline
\end{tabular}
\caption{Number of molecules and box geometries for the different sets of simulations for the calculation of the computational gain of water--ideal gas simulations compared to full atomistic and water--IBI simulations.}
\label{tab_speedup}
\end{center}
\end{table}

To avoid the inclusion of implementation and platform dependent run times we considered, in the measurements reported, only those quantities that differ for all-atom and adaptive H-AdResS simulations. Specifically, we only measured the time that the code spends with the calculation of non-bonded, pairwise forces and with the calculation of the drift term. In order to obtain results independent of parallelization, we ran the simulations on a single CPU. In parallelized adaptive resolution simulations, a load balancing protocol is required to benefit from the computational efficiency of coarse-grained potentials in the low resolution region.

\begin{figure}
	\includegraphics[width=\columnwidth]{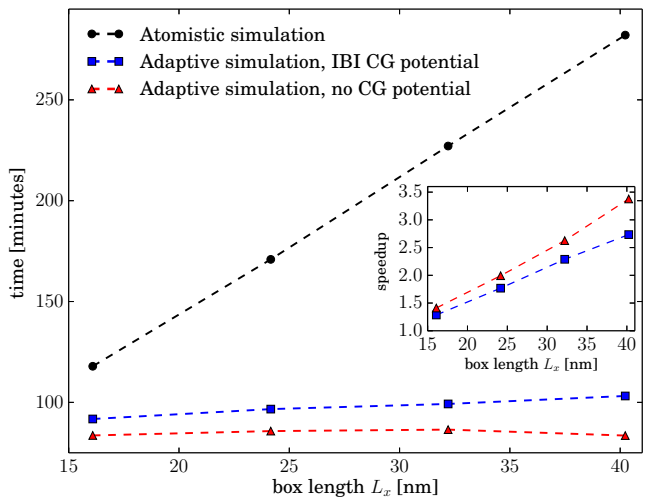} 
\caption{Main figure: times required for the non-bonded force calculations in the atomistic and adaptive simulations for systems of four different box lengths $L_x$. For the atomistic simulations, these are only non-bonded, pairwise forces, while in the adaptive H-AdResS simulations, also the time required for the drift term calculation is considered. Inset: speedup of the adaptive over the atomistic simulations. The lines are a guide to the eye.}
\label{fig:speedup}       
\end{figure}

The results of the simulations are plotted in Fig.~\ref{fig:speedup}, with the corresponding speedups being presented as an inset in the same figure. The speedups, reported in the inset of in Fig.~\ref{fig:speedup}, are defined as $T_\text{atomistic}/T_\text{adaptive}$, with $T_\text{atomistic}$ being the time spent for force calculations in the atomistic simulations and $T_\text{adaptive}$ being the corresponding time in the adaptive simulations.

It can be seen that the force calculations in the adaptive simulations are significantly faster than their corresponding atomistic counterparts. For the largest box, the non-bonded force and drift term calculations in the water--ideal gas simulations are faster by a factor of $\approx3.5$ than the full atomistic simulations. The water--IBI simulations reach a similar, though lower, speed-up. Therefore, the computational efficiency gained by coupling water to an ideal gas is slightly higher than the one gained by coupling to a typical coarse-grained potential. Furthermore, our results show that the time required for the adaptive simulations employing the ideal gas stays nearly constant for the different box sizes. The reason for it is that the simulation time is dictated by the interactions in the AT and HY subdomains, whose size does not change.

\section{Conclusions} \label{concl}

In tackling a wide spectrum of challenging problems in soft matter physics, dual resolution simulation methods can represent an advantageous simulation strategy. In fact, they allow us to provide a relatively small system, described with a highly detailed but computationally intensive model, with an accurate thermodynamical environment at a limited cost in terms of simulation resources. In particular, when the focus is concentrated on the high resolution subsystem, and the realistic modeling of the coarse-grained domain is of no interest, it is natural to push the simplification of the latter to the maximum.

With this goal in mind, we carried out the present work in order to study if an ideal gas can be employed as a highly coarse-grained model for water in an adaptive resolution setup. Our results, obtained from different simulation schemes, show that the basic requirements are indeed satisfied: in the force-based as well as in the energy-based case, the properties of the system in the AT domain are compatible with those that one would measure in a similar subregion cut from a fully atomistic simulation.

The equation of state of an ideal gas substantially differs from that of a liquid at the same temperature and density. In order to attenuate the deviations from the reference densities observed in both setups we introduce compensating external fields, namely the TF and FEC terms. These fields level out thermodynamical differences between the atomistic water model and the ideal gas, thereby maintaining the density of the fluid at the reference value.

Two-body correlations and relative fluctuations in the number of particles, as measured in the AT domain, perfectly reproduce those measured in a fully atomistic simulation. The diffusion profiles as well indicate that the dynamics in the high resolution region is not affected by the presence of a super-coarse-grained reservoir. These results are obtained at a very small computational cost compared to a fully atomistic simulation of a system having equal size. Additionally, the computational gain does not only come from the shorter time required to perform an integration step compared to a fully atomistic simulation. In fact, a faster diffusion of the solvent accelerates the configurational sampling. Consequently, the simulation duration necessary to equilibrate time independent physical observables is reduced. 

Last but not least, the major advancement allowed by the coupling of an ideal gas to a molecular fluid is given by the freedom to ``tune'' the physics in the low resolution region. The absence of any interaction in the latter enables one to easily insert and delete particles, making it possible to regulate thermodynamical quantities such as pressure and chemical potential, and employ smaller simulation boxes without introducing finite size effects.

Some issues, however, still remain open. For example, the mean field character of the FEC obtained through Kirkwood TI makes it insufficient, in the H-AdResS simulations, to completely remove a small but noticeable deviation from the reference density at the HY/CG interface. This difficulty might prove to be particularly challenging when the focus moves towards more complex systems, such as solvent/cosolvent mixtures. Specifically for this case, in fact, it might be worth employing more accurate, iterative algorithms to construct an external field which counteracts the average drift force exactly, as suggested in Ref. \cite{JCPhadress2014}. A further possibility could be to combine different methods, namely the FEC and the TF.

\section*{Acknowledgments}

K. Kreis is recipient of a fellowship funded through the Excellence Initiative (DFG/GSC 266). K. Kremer and A. Fogarty acknowledge research funding through the European Research Council under the European Union's Seventh Framework Programme (FP7/2007-2013) / ERC grant agreement n. 340906-MOLPROCOMP. R. Potestio and K. Kremer acknowledge funding from the SFB--TRR 146 grant. The authors are indebted to T. Bereau and D. Mukherji for a careful reading of the manuscript.


\providecommand*{\mcitethebibliography}{\thebibliography}
\csname @ifundefined\endcsname{endmcitethebibliography}
{\let\endmcitethebibliography\endthebibliography}{}

\end{document}